# TOWARDS TECHNOLOGY INDEPENDENT STRATEGIES FOR SOA IMPLEMENTATIONS

Zheng Li
*School of Computer Science, ANU and NICTA, Canberra, Australia*
*Zheng.Li@nicta.com.au*

He Zhang
*School of Computer Science and Engineering, UNSW and NICTA, Sydney, Australia*
*He.Zhang@nicta.com.au*

Liam O'Brien
*School of Computer Science, ANU and CSIRO, Canberra, Australia*
*Liam.OBrien@csiro.au*

Abstract: Benefiting from the technology based strategies, Service-Oriented Architecture (SOA) has been able to achieve the general goals such as agility, flexibility, reusability and efficiency. Nevertheless, technical conditions alone cannot guarantee successful SOA implementations. As a valuable and necessary supplement, the space of technology independent strategies should also be explored. Through treating SOA system as an instance of organization and identifying the common ground on the similar process of SOA implementation and organization design, this paper uses existing work in organization theory area to inspire the research into technology independent strategies of SOA implementation. As a result, four preliminary strategies that can be applied to organizational area we identify to support SOA implementations. Furthermore, the novel methodology of investigating technology independent strategies for implementing SOA is revealed, which encourages interdisciplinary research across service-oriented computing and organization theory.

## 1 INTRODUCTION

Service-Oriented Architecture (SOA) emerges with the requirements of quick response to the rapid and often unpredictable changes in business environment for modern enterprises. Motivated by the expectations of the people who are engaged in SOA activities, SOA has goals such as reusability of software assets across multiple platforms and applications, agility of support to business processes, efficiency in terms of development time and cost, and flexible integration of existing and legacy information systems. Considering *goals* define where we want to go while *strategies* define how we will get there (Daft, 2009), SOA implementation strategies should be investigated to facilitate achieving SOA's goals. In fact, numerous strategies have been proposed and developed over the past decade. For example, Newcomer and Lomow (2004) use Web services to concrete the conceptual SOA; Krafzig, Banke and Slama (2004) focus on the practical application of SOA in enterprises with discussion of the roadmap of relevant technologies; Erl (2007; 2009) summarizes a full scope of implementation strategies through eight principles of service design and 17 SOA design patterns. To the best of our knowledge, however, most of existing strategies for SOA implementation only pay attention on the technology aspect, particularly relying on the current state of the art of Web technology. As we know, technology is a necessary but not a sufficient condition for successful SOA implementations (Rosen, Lublinsky, Smith and Balcer, 2008). In other words, technology cannot guarantee the success of an SOA system. Therefore, the technology independent strategies for implementing SOA should also be identified and examined.

To achieve technology independent strategies of SOA implementation, it would be necessary to investigate every aspect of SOA system. In general, "SOA is a concept for large distributed systems" (Josuttis, 2007), which supposes services are decentralized and may be under the control of different owners. When unfolding research into distributed systems, we can generally adopt two different approaches (Fox, 1981): one is learning by doing (building real distributed systems), while the other is learning by analogy (drawing upon ideas from other research areas). Considering the limitations to empirically implementing various SOA projects by ourselves, we have employed learning by analogy with organization theory as an efficient way to inspire the research into SOA domain. In fact, the use of organizational theory to guide technology research has proven significantly beneficial particularly in the multi-agent system community. For example:

- Well-known human organizational structures are used for the deployment of multi-agent systems (Argente, Julian and Botti, 2006);
- Social laws are chosen to simplify multi-agent systems (Fitoussi and Tennenholtz, 2000);
- Dependency theory of social interaction is used to explain how to achieve social goals of multi-agent systems (Sichman and Demazeau, 2001).

Therefore, this paper also presents an organization-based view to comprehend SOA, and treats SOA implementations as organizational activities. Based on the traditional consensus of the organization concept, thinking of SOA organizationally could be reasonable. Moreover, the parallels are identified between organization design and SOA implementation in general, which follows a pentagonal process with five steps focusing on the *Goal and Strategy*, *Environment and Scope*, *Structure*, *Process* and *Coordination and Control*. Note that the "SOA implementation" we discuss here refers to common SOA implementation practices rather than any particular case. Enlightened by existing work of organization design in the organization theory domain, we have initially identified four strategies as a demonstration to meet four predetermined research topics of service-oriented software engineering (Kontogiannis, Lewis and Smith, 2008): (1) to use Total Quality Management (TQM) to accommodate the Quality Assurance challenge in the Engineering topic, (2) to keep the structure as flat as possible meet the Adoption challenge in the Operations topic, (3) to take measurements at interim steps in process to satisfy the Governance challenge in the Cross-Cutting topic, and (4) to build business process teams to facilitate mapping between business structure and service-oriented environment in the Business topic. These four strategies are independent and each can improve SOA implementations depending on real circumstances. In other words, the strategies can be employed both individually and all together for an SOA implementation instance.

Overall, this paper makes three contributions. Firstly, new interdisciplinary research opportunities are suggested between service-oriented computing and organization theory. Secondly, the methodology of investigating technology independent strategies for SOA implementation is outlined. Thirdly, four preliminary strategies enlightened by organization design are identified within four predetermined research topics of SOA.

The remainder of the paper is organized as follows. Section 2 justifies thinking of SOA from an organizational perspective. Section 3 analogizes the procedure of SOA implementation with that of organization design. Section 4 introduces four technology independent strategies enlightened by organization design for SOA implementation. Section 5 uses an example to demonstrate how these four strategies are applied to improve SOA implementation. Conclusions are drawn and some future work is proposed in Section 6.

## 2 SOA: AN ORGANIZATIONAL PERSPECTIVE

Organizations emerged as early as ancient civilizations appeared. Today, organizations have become indispensable and pervasive components of human beings' society, for example, from schools to hospitals and from armies to governments. When it comes to the SOA area, we can similarly regard service-oriented systems as virtual organizations that are composed of services. There are two ways of thinking of SOA from the organizational perspective. One is to view an SOA system as an organization; the other is to treat SOA system as a mirror of its corresponding organization.

### 2.1 Viewing an SOA System as an Organization

Viewing SOA systems as organizations is to use the organization concept to cover SOA systems, as shown in Figure 1. Under the same umbrella of organization concept, both traditional organizations and SOA systems consist of organizational units. Organizational units in an SOA system are services, while that in a traditional organization are individuals. Furthermore, different organizational units have different skills and play different roles in an organization. For example, composite services play integrative roles in an SOA system, which parallels the responsibilities of managers in a traditional organization. Unfortunately, there is no single agreement on definition of an organization. Fierce debates about the organization concept are still underway, though theorists have traditionally consented that organizations are collectivities of people who are socially arranged to pursue specific purposes and achieve explicit goals (McAuley, Duberley and Johnson, 2007). This classical consensus makes it possible to think of SOA from the organizational perspective due to two reasons.

First, it is suitable to think of SOA representing organization architecture. The Organization for the Advancement of Structured Information Standards (OASIS) (2006) defines SOA as "a paradigm for organizing and utilizing distributed capabilities that may be under the control of different ownership domains. It provides a uniform means to offer, discover, interact with and use capabilities to produce desired effects consistent with measurable preconditions and expectations." When it comes to implementation, SOA is used to build up a collection of independent services that can be quickly and easily integrated into different, high-level business services and business processes to create business value and achieve business strategies (Rosen et al., 2008). To summarize, SOA both in theory and in practice is proposed for organizing services to attain some particular goals. Therefore, SOA can be set under the umbrella of organization theory in terms of the suggestion of traditional organization concept: if the organizing process is about goal attainment, the organization theory could be followed to conceptualize, explain and ultimately guide individuals' activities that should be united together to achieve desirable, common organizational goals (McAuley et al., 2007).

On the other hand, it is reassuring to think of SOA from the organizational perspective. In fact, conceptual challenge might appear when talking of organizations based on having a goal, because the agreement about an organization's purpose amongst members may not exist. In the SOA area, however, this disagreement issue can be ignored. Within SOA systems, a service is a well-defined unit of functionality realized by a service interface and a service implementation (Papazoglou and Heuvel, 2007). A service interface identifies a service and exposes the semantic description of the service's invocation. A service implementation realizes the work that the service is designed to perform. Unlike people in social organizations, services in SOA do not have mental or psychological attributes. Consequently, services will always obey the control from the "senior management" of the whole SOA system, and may even not be aware of the "organizational goal". When thinking about SOA organizationally, the blind obedience characteristic of services can naturally avoid the challenge of defining organizations in terms of having a goal while not all members freely agree to that goal (Bakan, 2005).

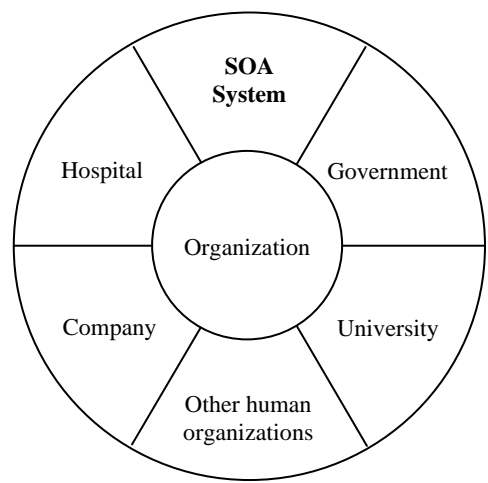

Figure 1: View SOA system as an instance of organization.

Moreover, according to a set of general characteristics of the organizations identified by Campbell and Craig (2005), we can find more similar features between SOA system and organization. For example:
- Both human organization and SOA system contain organizational units (people/services);
- These organizational units perform different roles/functionalities and their employment depends on such performance;
- Any human organization or SOA system has a collective goal to which all units subscribe;
- All of the roles/functionalities, taken together, help the human organization / SOA system achieve its collective goal;
- Difference tasks are distributed to different individual units according to their expertise,

interest or specialism (human) / functional capability and non-functional performance (service);
- There is a clearly defined hierarchy of authority so that each member of the organization is aware of where he or she 'fits in'. / This scenario depends on Orchestration or Choreography when composing services in SOA system;
- The limits or borders of a human organization / SOA system are usually clearly defined.

## 2.2 Treating an SOA System as a Mirror of Real Organization

Unlike Conway's Law (Conway, 1968), treating SOA systems as copies of real organizations herein implies that an SOA system reflects the features of an organization for which the system is implemented. Conway's Law emphasizes the organization who designs a system, which states that "any organization that designs a system will inevitably produce a design whose structure is a copy of the organization's communication structure" (Conway, 1968). However, here we emphasize the organization for which a system is designed and implemented. Figure 2 illustrates such a sample that a departmentalized company and its SOA based information infrastructure have the same functional structures.

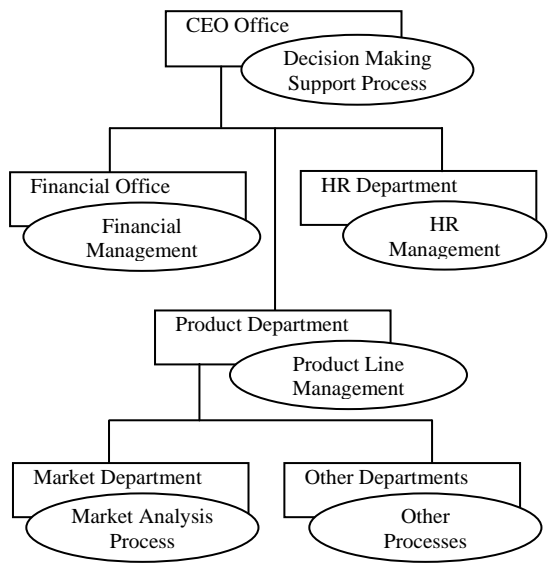

Figure 2: A Company and its infrastructural SOA system.

In fact, current information infrastructures for organizations are normally realized by third-party software companies, but the organizations themselves are also closely involved in the activities from requirements analyses to system implementations. Considering the information technology (IT) must be aligned with business when establishing information infrastructure for an organization, any SOA based information system will inevitably copy the features of the organization for which the SOA system is designed and built. For example, services will be grouped functionally to support different departments in a functional structure based organization, while services should be grouped separately according to different product lines in an organization that employs product-based structure; businesses with narrow span of control will result in tall-hierarchy control flows among services, while businesses adopting broad span of control will bring flat-hierarchy control flows among services.

Benefitting from thinking of SOA from the organizational perspective, no matter treating SOA system as a mirror of real organization or viewing SOA system as an organization, we can use organization concept to comprehend SOA systems, and further use organization theory to inspire SOA implementations.

## 3 ANALOGIES BETWEEN SOA IMPLEMENTATION AND ORGANIZATION DESIGN

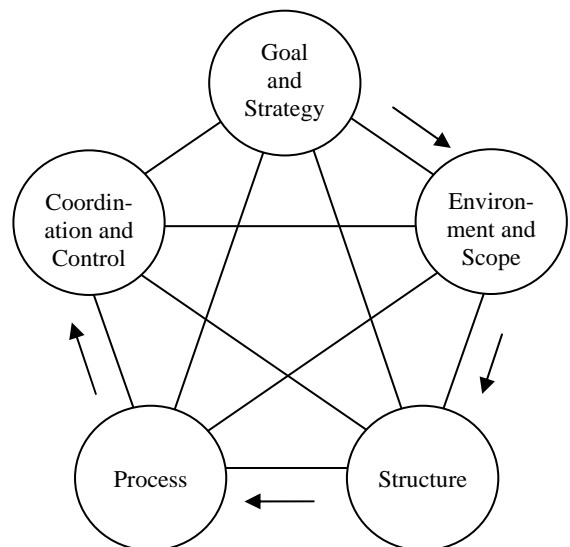

Figure 3: The pentagonal process of SOA implementation / organization design.

The similarity between SOA systems and organizations is not a coincidence. We can find common ground on a pentagonal process of SOA implementation (Josuttis, 2007; Krafzig, 2004; Lawler and Howell-Barber, 2007; Rosen et al., 2008) and organization design (Burton, DeSanctis and Obel, 2006; Daft, 2009; Davis and Weckler 1996; Kates and Galbraith, 2007), which is identified through refining the waterfall process of organization design (Burton et al., 2006). In the pentagonal process, five steps focusing on the *Goal and Strategy*, *Environment and Scope*, *Structure*, *Process* and *Coordination and Control* are executed generally along the clockwise sequence arrowed in Figure 3. Meanwhile, each step has influence on as well as is under influence of the other four steps. For example, the goal and strategy together determine the whole process of organization design or SOA implementation, while they will be refined gradually as the process is unfolded.

### 3.1 Goal and Strategy

As mentioned previously, an organization must have a collective goal according to the traditional consensus of organization concept. Although different parts of the organization may have their own objectives, an overall collective goal can be established by aggregating all the separated objectives together. The overall goal is a desired direction that the organization will head. In practice, an organization's overall goal embodies a set of specified goals, each of which focuses on different aspect of the organization. Daft (2009) distinguishes organization's goals into official goals and operative goals. The official goals formally define the business, values and outcomes that the organization attempts to achieve, while the operative goals are more explicit and scattered in different facets such as performance, efficiency, innovation and profit.

Goals of an organization introduce the target that the organization wants to pursue, while strategies define how the organization can pursue its target. Therefore, strategies can be treated as the operationalization of organization's goals (Burton et al., 2006). Following the analysis of organization's goals, we can also distinguish organization's strategies into official strategies and operative strategies. The official strategies are essential plans of actions that can realize the corresponding official goals, for example the cost-leadership strategy or differentiation strategy. On the other hand, the operative strategies will aim at different detailed tasks like how to improve working efficiency or increase product profits. As different tasks may have resource conflict with each other, the strategy set should be carefully balanced.

Goals and strategies are in the first phase of organization design and essentially influence how an organization should be designed. Similarly, the first step of SOA implementation is to identify the business strategies and goals, and we can adopt the technique, namely business value chain, to help identify the specific goals and strategies for certain SOA projects (Rosen et al. 2008). Since service-oriented computing emerged from the requirement of addressing the rapid and usually unpredictable changes that modern enterprises are confronting, SOA systems contribute more promises than the traditional software infrastructures. Therefore, common goals and strategies can be extracted among the general SOA implementations, which are emphasized in Section 4.

### 3.2 Environment and Scope

The environment is the surroundings of a system, and the system influences and is influenced by its environment. Meanwhile, the environment is not static but can be changing continuously and dynamically. Generally, there are five environment patterns interacting with any system, including asymptotic variation, interfering variation, periodic variation, phase-transition variation, and random variation (Peng, Liu and Tao, 2009).

Both SOA systems and organizations cannot be isolated from their external environments. The environment surrounding an SOA system or organization has a set of factors relating to resources or vulnerabilities. For example, the suppliers, customers, competitors, culture and government are organizations' environmental factors, while the developers, users, legacy system, existing service pool and state of current technology are SOA systems'. Building organization and implementing SOA are highly dependent on the environmental factors. In practice, the number of factors that constitute environment might be considerable. All these factors together reflect the boundary that an organization or SOA system, and then outline a scope, which determines the capability, applicability, competitive advantages and business range for the organization or SOA system.

For organization design, environment restricts organizations within certain scopes, and further influences their processes, structures and controls. For SOA implementation, analyzing the external environment and determining the applicable scope are particularly significant. SOA-based software infrastructure is supposed to be adaptive within an increasingly changing and complex environment. However, the loosely coupled asynchronous SOA systems are inherently more complex than the traditional architecture based systems. Josuttis (2007) has pointed out that distributed processing would be inevitably more complicated than non-distributed processing, and any form of loose coupling increases complexity. In practice, building a true heterogeneous SOA for a wide range of operating environments may take years of development time if the company does not have sufficient SOA experience and expertise (Jamil, 2009). Since the more complexity involved in a system, the more difficulty the designers or engineers have to understand the implementation process and thus the system itself (Cardoso, 2005), SOA should be adopted only in the suitable environment and only when its benefits outweigh any extra costs due to the increased complexity.

### 3.3 Structure

Structures of both organizations and SOA systems are established to divide up the work into manageable and measureable units with clear responsibility boundaries. Organization's structure is normally a hierarchy that allocates roles, power, authorities and responsibilities, and determines working relationships and communication channels. Generally, organizational units are arranged around functions, products/services, customers/geographies, or business processes. Therefore, the organization structures can be typically divided in five basic styles: functional, product- or service-based, customer- or geographical area based, business process based, and matrix structure (Davis and Weckler, 1996). Each kind of structure has specific advantages and disadvantages. Unsuitable organization structure will result in formidable obstacle to align the other design elements with the organization's strategy (Kates and Galbraith, 2007). Consequently, large organizations always build hybrid structures to achieve the combination of the advantages.

SOA systems normally adopt matrix structure, which simultaneously groups services in two directions: functional direction and business process based direction, as shown in Figure 4. The functional direction is to classify services according to the type of logic they encapsulate. Although there are quite a few service classifications we can identify from literature, most of the existing classifications can be unified and layered as Basic Services, Business Entity Services, Process-centric Services, and Enterprise Services. The Basic Services, settled at the bottom layer of SOA systems, provide reusable, technical, and foundational functionalities. The Business Entity Services represent the entities in business activities, such as employee, customer, contract, and product. Through composing relevant Business Entity Services, a Process-centric Service encapsulates a sequence of activities to complete a specific business task. The Enterprise Services provide endpoints to access the corresponding SOA systems, which could have less reuse potential but enable cross-enterprise integrations.

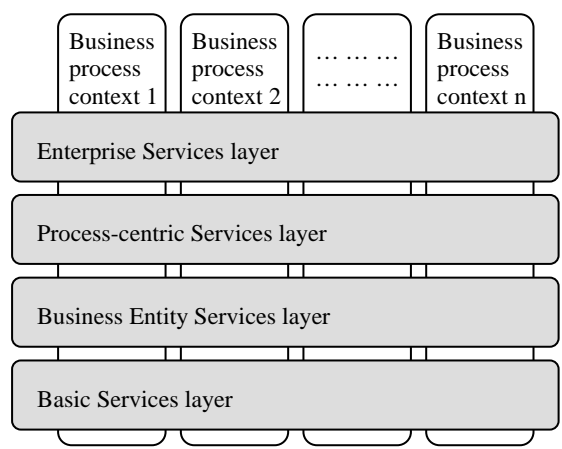

Figure 4: The matrix structure of SOA systems.

On the other hand, to obtain the reuse of services in multiple business processes, the landscape of different services within different business process contexts should be described to show how the services work together (Rosen et al., 2008). The business process based direction of SOA systems' structure can then be outlined. Along the business process based direction, the overall services are grouped according to different roles and responsibilities within the real business. Each group may contain single service or multiple services. Moreover, the relationships to each other among the groups and the places of these groups in the business processes are also described.

### 3.4 Process

The organization design has been viewed from an information processing perspective (Galbraith, 1974). The term process herein means not only business processes that produce products or services for customers, but also non-profit routines that constitute organizational actions. Generally, a process in an organization is a series of connected activities that transfer and transform information and resources through the organization (Kates and Galbraith, 2007), for example approving an application, submitting a report, and managing work progress. From the mid-1990s, many modern enterprises began to evolve into process-focused organizations in order to achieve higher performance and survive the market competition (Seltsikas, 2001). Currently, process-focused organization has become the new organizational form with business process as the core concern. As a result, the design of processes significantly impacts on how well the organizational goals can be achieved.

Moreover, processes in an organization have close relationships with the organization's structure and coordination. All kinds of organization's structures inevitably create barriers to collaboration, because boundaries will appear as soon as organizational units are grouped under the structure. To fulfil the effective collaboration targets in organization, however, processes are required to flow cross the boundaries. Therefore, processes can glue the related organizational units to work together.

Like process-focused organizations, SOA systems can also be regarded as process-focused systems comprising such as management process, coordination process, and traditional work process. Among all kinds of processes, SOA inherently concentrates on business process. Essentially, SOA is aligned with business process management (BPM) in business firms in which the criticality of business processes is concerned (Lawler and Howell-Barber, 2007). The emphasis of SOA is the functional infrastructure and the business services instead of the technical infrastructure and technical services. A business service encapsulates a piece or an entity of a business process. When implementing SOA, it is crucial to analyze business processes before identifying and developing services (Rosen et al., 2008). Following the analysis of business process, those potentially and even partially suitable services should be identified first. These existing services provide constraints that frame the future SOA system. The business processes are then broken down into business pieces that can be implemented by developing new services.

### 3.5 Coordination and Control

The coordination problem is one of the central topics in organizational studies (Heath and Staudenmayer, 2000). As mentioned previously, individual actions of large numbers of interdependent roles and specialists must be coordinated to constitute processes to fulfill global tasks in an organization. On the other hand, the coordination will increase organizations' information processing capabilities when encountering increasing amount of uncertainty (Galbraith, 1974). In practice, the activity of coordinating overlaps the activity of controlling, because the appearance of coordination usually implies the occurrence of some control (Davis and Weckler, 1996). To coordinate and control organizational work, organizations should adopt suitable techniques and mechanisms. Unfortunately, there is not a fixed prescription of methods for coordinating and controlling work. The coordination and control, for example, can be simply related to the structures (Kates and Galbraith, 2007), be utilized by goal setting, hierarchy, and rules (Galbraith, 1974), or be executed by using four basic techniques: Supervision, Standardization, Building employee commitment, and Teams (Davis and Weckler, 1996). However, the principle of these techniques and mechanisms is uniform: to make sure organizational units work appropriately and find out to what extent they are reaching the goals and targets.

When implementing SOA, services must be composed to fully realize the benefits of SOA (Sarang, Jennings, Juric and Loganathan, 2007), which also relies on the coordinating and controlling activities. According to the cooperation fashions among component services, the mechanism of coordination and control can be distinguished between *Orchestration* and *Choreography*. Orchestration describes and executes a centralized process flow that normally acts as an intermediary to the involved services. Choreography describes multi-party collaboration and focuses on the peer-to-peer message exchange.

If comparing Orchestration and Choreography with the two classical organization types – Mechanical and Organic organizations, we can find that the fundamental ideas and notions behind these different concepts in two disciplines are nearly the same. In particular, we can even explain Orchestration and Choreography by using the descriptions of those two types of organizations

## 4 TECHNOLOGY INDEPENDENT STRATEGIES FOR SOA IMPLEMENTATION

When it comes to information system architecture, service-orientation establishes a universal model in which functionalities and business logics are cleanly partitioned and consistently represented. Therefore, in addition to the various targets that satisfy the business requirements of respective SOA projects, SOA implementation possesses general promises and goals. The motivations and expectations of the people who are engaged in SOA activities can be used to empirically and efficiently assess the universal goals of SOA. With reference to the survey in 2006 conducted by the Cutter Consortium (Rosen et al., 2008), we can identify the most common and general goals of SOA are agility, flexibility, reuse, data rationalization, integration, and reduced costs. In fact, plenty of technical strategies have been developed to help realize SOA's goals. Examples of technology based strategies are standard service contract that facilitates integration, loosely coupling that supports flexibility, and service autonomy that provides reliable and predictable performance. As supplementary, here we focus on the technology independent strategies of SOA implementation that are inspired by existing research into organization design. Following four research topics of service-oriented software engineering (Kontogiannis et al., 2008), we have initially identified four strategies respectively (S1~S4).

### 4.1 S1: Applying TQM to SOA Implementation

Under the Engineering research topic of service-oriented computing, we propose to use Total Quality Management (TQM) to accommodate to the quality assurance of SOA implementation. Just as the name implies, TQM is a holistic level management for quality, because it can be achieved only if the total quality concept is utilized from the acquisition of resources to the customer satisfaction (Kaynak, 2003). When it comes to SOA, the quality management has been emphasized to satisfy the unique characteristics of service-oriented computing. Nevertheless, to the best of our knowledge, existing research into quality management in SOA area is mainly at the service level, which is limited around the Quality of Service (QoS). The overall QoS of an SOA system is determined by all the QoS of component services who compose the SOA system (Yau, Ye, Sarjoughian and Huang, 2008). Based on the QoS management, SOA systems generally replace component services with higher quality services to realize adaptations. Hence, the focus of QoS management is on individual services in an SOA environment.

When applying TQM to SOA domain, Deming's 14 points (Walton, 1988) can be used as a framework to guide SOA implementations. For example, service suppliers and SOA system users should be taken into account when measuring the total quality of an SOA implementation. Here we focus on the quality of interaction and cooperation process among services. With reference to the explanation of TQM by Deming, in any circumstance, processes should be constantly analyzed to determine what changes can be made to bring improvement. Therefore, TQM introduces a new angle of view to SOA systems when adapting environment. However, employing TQM does not indicate abandoning QoS management. There is no conflict between TQM and QoS management. On the contrary, they are two complementary approaches for SOA to accommodate the changing environment: (1) TQM can be used to adjust the process of interaction and cooperation among services. (2) QoS management can be used to switch services based on the latest quality requirement.

### 4.2 S2: Flattening the Structure of SOA Systems

Under the Operations research topic of service-oriented computing, we propose to flatten the structure of SOA systems when considering service composition. In human organizations, every level in a hierarchy will inevitably involve more operating costs (George and Jones, 2007). An organization with a higher hierarchy may come with longer decision making chains and slower responsiveness to customers. Therefore, organizations can increase efficiency by keeping their structures as flat as possible. Furthermore, flat structure can decentralize responsibility and control to lower-level employees to take greater advantage of the skills and experience of organization members.

In an SOA system's hierarchy, the number of levels can increase along with the growing cascade of service composition. In general, a composite service is recursively defined as an aggregation of elementary and composite services. When thinking of SOA from the organizational perspective, composite services play integrative roles in an SOA system. In organizations, an integrative role is a full-time manager who is in charge of orchestrating work across units (Kates and Galbraith, 2007). These managers have accountability for results but are not directly responsible for the resource achieving and specific work that should be accomplished by staff. A flat organizational structure can help reduce the number of integrative roles. The similar scenario of SOA implementation can be simply illustrated in Figure 5.

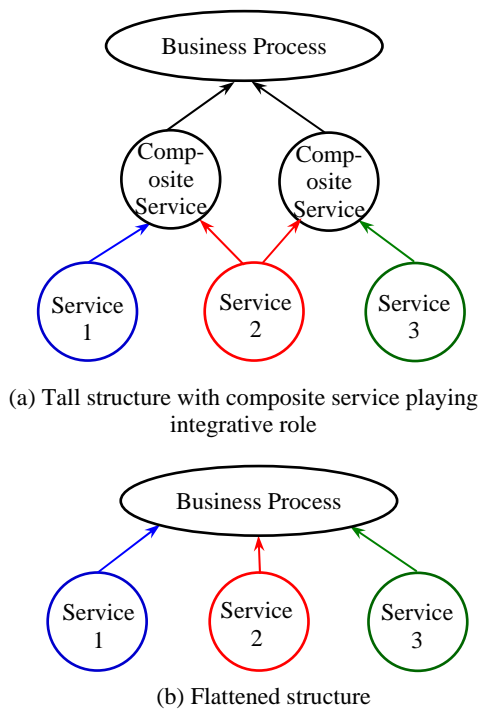

(a) Tall structure with composite service playing integrative role

(b) Flattened structure

Figure 5: Tall structure vs. flat structure for implementing the same business process.

Figure 5(a) shows a tall structure example of some business process implementation by inserting two composite services that respectively compose two of the elementary services. Suppose only the three elementary services are available at the bottom, we can flatten the structure and move the additional functionality of original composite services upward, as shown in Figure 5(b), to reduce the composition cost and lower the complexity of the business process implementation. However, we should keep the tall structure if the composite service already exists or its reusability is to be achieved. Therefore, when applying this strategy, the value and cost should be well balanced to determine the extent of flattening structure.

### 4.3 S3: Taking Measurements at Interim Steps in Process

Under the Cross-Cutting research topic of service-oriented computing, we propose to take measurements at interim steps in process to govern SOA implementation. When generating products following certain working processes or designing the working processes in an organization, it has been proven valuable to take measurements at interim steps in processes. The research and practice in organizations during the past decades reveal that it is increasingly important to ensure the work finishes properly the first time instead of having to be redone (Davis and Weckler, 1996). The inspections and measurements can be applied to different steps in processes to save the cost of rework and avoid flaws in the end product.

When applying this strategy to SOA implementation, the inspiration is to confirm the individual work of each service in processes. The idea behind this strategy is to clearly define connected subtasks in a process, and specify and measure the result of each subtask. It should be noted that measuring interim task mainly concerns the result rather than how the task is performed. Considering a service is such an entity that performs some task while hiding technical details, we can use the interim task measurement to help identify the most suitable services. Once all the services are determined, the relevant business process can then be correctly implemented.

### 4.4 S4: Building Virtual Business Process Teams

Under the Business research topic of service-oriented computing, we propose to build business process teams to facilitate mapping between business structure and service-oriented environment. In human organizations, teams are cross-functional structures that bring people outside the scope of traditional departments to work together and share collective responsibility for special and complex assignments. A business process team is established around one business process and includes people who can collectively perform all the major activities to carry out the business process from beginning to end. Business process teams exist as a characteristic of "horizontal organizations" (Davis and Weckler, 1996). Horizontal organizations instinctively tend to flatten their structures by focusing on all the units involved in completing certain work, rather than the coordinate activities relying on a vertical hierarchy. As analyzed in Section 4.1, horizontally established business process teams then have advantages of reduced management costs and less need for coordination.

Building business process teams in an SOA system should be a virtual division without many real actions. All the services involved in a business process logically constitute a team without changing the existing structure of the SOA system. Through virtual business process teams, the focus of coordination and control can be balanced between inward IT and outward business during SOA implementations. Furthermore, considering one service can be involved in different business process teams like the same scenario of organizational teams, we can identify and scale services' dependency of business processes in an SOA system. The more dependency a service has, the more carefully it should be controlled especially when planning to modify or adapt this service.

# 5 AN EXAMPLE CASE

Here we use a simplified case to demonstrate how the technology independent strategies can be applied to improve an SOA implementation. The example case is an SOA-based application in a travel agency, as illustrated in Figure 6.

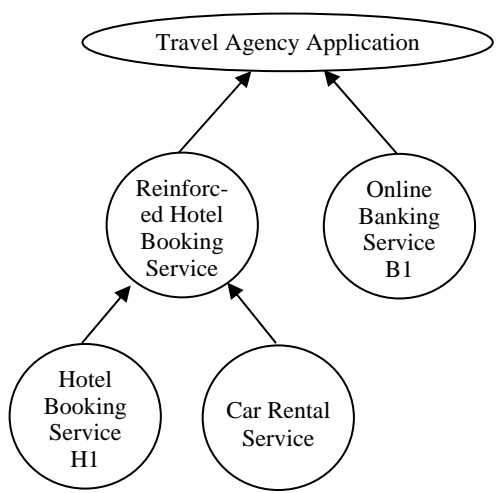

Figure 6: An SOA-based application of hotel booking in a travel agency.

The travel agency books hotel through BPay on behalf of a group of tourists, and will rent a car by money transfer if the number of the tourists is more than ten. Suppose there are three online banking services, two hotel booking services and one car rental service. Each service can fulfill its corresponding business function, while Online Banking Service B1 and Hotel Booking Service H1 are selected according to their reliability and response time. Moreover, the business rule "*rent a car by money transfer if the number of the tourists is more than ten*" is encapsulated in a composite service composed by Hotel Booking Service H1 and Car Renting Service. The composite service is implemented as a Reinforced Hotel Booking Service following the technology based strategy of using service composition to "fulfill a large extent of future business automation requirements" (Erl, 2007). After applying the four proposed strategies, this travel agency application will evolve as shown in Figure 7.

## 5.1 Applying TQM (S1)

After a period of operation, the travel agency receives many complaints from small groups of tourists about inconvenience without cars. Hence, the travel agency decides to change the business rule into "*rent a car by money transfer if the number of the tourists is more than five*". When applying TQM to the SOA system to check the cooperation among services, we can find that the invocation of Car Rental Service is inflexible because old business rule is hardcoded in the Reinforced Hotel Booking Service. Therefore, the number of tourists should be set as a variable and exposed as an input parameter of the composite service. The operation of this composite service is then adjusted by accepting one threshold parameter to improve the flexibility of the SOA system.

## 5.2 Flattening the Structure (S2)

When analyzing the structure of the travel agency's SOA system, we find that the Reinforced Hotel Booking Service does not have any reuse opportunity. Furthermore, the encapsulated business rule can be easily transformed into control flow logic of invoking two component services, and moving the control flow logic into upper business logic will have little increase complexity for the latter. Therefore, we can flatten the structure as Figure 5(b) by removing the Reinforced Hotel Booking Service to reduce the service maintenance effort.

## 5.3 Measuring Interim Steps in Process (S3)

Suppose both BPay and money transfer will result in commission charges, and the charges vary depending on different bank and transaction time. We can then use the criterion "*choose bank with the lowest commission charges*" to constantly and simultaneously measure the three candidate online banking services. The service of the bank that charges the lowest fee will be dynamically employed by the SOA system to help the travel agency save money.

### 5.4 Building Virtual Teams (S4)

Based on the business logic behind the travel agency application, we can identify there are two atomic business processes: one is hotel booking through BPay, and another is car rental through money transfer. Consequently, the selected Online Banking Service, Hotel Booking Service and Car Rental Service can be logically grouped into two business process teams. The coordination and control among services in the hotel booking business process team follows the BPay rules, while in the car rental business process team obeys the money transfer rules. Through the team building, we can naturally arrange different cooperation for services in different teams, and in this case we may further identify the Online Banking Service is the key service when implementing the SOA system.

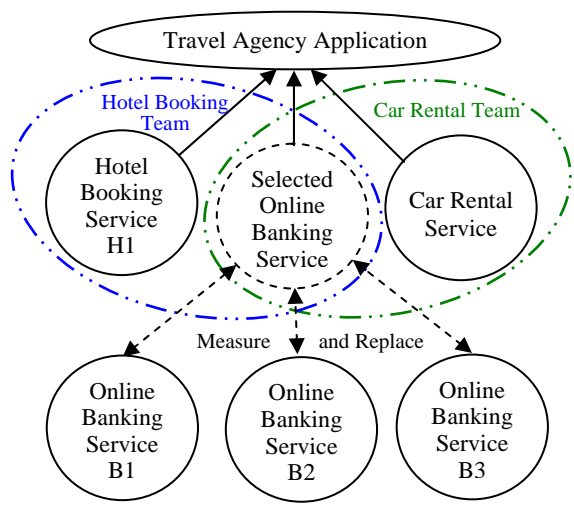

Figure 7: The travel agency application after applying four proposed strategies.

## 6 CONCLUSIONS

The emergence of SOA has been considered a feasible opportunity for modern enterprises to leverage the capabilities of quickly adapting to competitive and changing environment. Compared with systems based on traditional architecture, however, SOA systems are inherently more complicated. Since the more complexity involved in a system, the more difficulty the designers or engineers have to understand the implementation process and thus the system itself (Cardoso, 2005), it is vital to find a set of implementation strategies to help achieve the promises of SOA. Based on the review of the relevant literature, we can identify a suite of technical strategies based on the current state of the art of Web technology. For example, to hide the technical details to abstract services, to standardize the contract of communication between services, to decouple the logic as well as the implementation of services, and to use service bus to facilitate service integration. Whereas, technology based strategies cannot guarantee the success of SOA implementations (Rosen et al., 2008). We therefore notice that the technology independent strategies should also be emphasized as the supplements in implementing SOA.

By presenting an organization-based view to comprehend SOA, and treating SOA implementations as organizational activities, this paper delivers three main contributions. First, interdisciplinary research opportunities are suggested across SOA area and organization theory area. Second, the methodology of investigating strategies for implementing SOA is proposed by analogizing organization design with SOA implementation. Last, benefiting from existing work of organization design in the organization theory domain, four preliminary strategies conforming to the general goals of SOA are identified and suggested at this stage.

The four technology independent strategies highlighted in this paper can be applied to different steps in the process of organizational design. When applying them in an SOA environment, one simplified example case is elaborated to show their applicability for SOA implementations. However, their value and effectiveness still need to be further investigated and evaluated in practice. Therefore, the future work of this research is to explore other technology independent strategies for SOA implementation, as well as to apply these strategies in real scenarios and report the empirical results.